\documentclass[conference]{IEEEtran}
\def \toolname {\textit{IncBL} }

\IEEEoverridecommandlockouts
\usepackage{cite}
\usepackage{amsmath,amssymb,amsfonts}
\usepackage{algorithmic}
\usepackage{graphicx}
\usepackage{textcomp}
\usepackage{xcolor}
\usepackage{url}

\usepackage{comment}
\newboolean{showcomments}
\setboolean{showcomments}{true}
\ifthenelse{\boolean{showcomments}}
{ }

\ifthenelse{\boolean{showcomments}}
{ }

\ifthenelse{\boolean{showcomments}}
{ }

\def\BibTeX{{\rm B\kern-.05em{\sc i\kern-.025em b}\kern-.08em
    T\kern-.1667em\lower.7ex\hbox{E}\kern-.125emX}}
\begin{document}

\title{IncBL: Incremental Bug Localization}

\author{
\IEEEauthorblockN{Zhou Yang\textsuperscript{$\ast$}, Jieke Shi\textsuperscript{$\ast$} \thanks{$\ast$ Equal contributions.}}
\IEEEauthorblockA{
\textit{Singapore Management University}\\
Singapore \\
\{zyang, jiekeshi\}@smu.edu.sg}
\and
\IEEEauthorblockN{Shaowei Wang}
\IEEEauthorblockA{
\textit{University of Manitoba}\\
Canada \\
shaowei@cs.umanitoba.ca}
\and
\IEEEauthorblockN{David Lo}
\IEEEauthorblockA{
\textit{Singapore Management University}\\
Singapore \\
davidlo@smu.edu.sg}
}


\maketitle

\begin{abstract}	Numerous efforts have been invested in improving the effectiveness of bug localization techniques, whereas little attention is paid to making these tools run more efficiently in continuously evolving software repositories. This paper first analyzes the information retrieval model behind a classic bug localization tool, BugLocator, and builds a mathematical foundation illustrating that the model can be updated incrementally when codebase or bug reports evolve. Then, we present \textit{IncBL}, a tool for \textbf{Inc}remental \textbf{B}ug \textbf{L}ocalization in evolving software repositories. \toolname is evaluated on the Bugzbook dataset, and the results show that \toolname can significantly reduce the running time by 77.79\% on average compared with the re-computing the model, while maintaining the same level of accuracy. We also implement \toolname as a Github App that can be easily integrated into open-source projects on GitHub. Users can deploy and use \toolname locally as well. The demo video for \toolname can be viewed at \url{https://youtu.be/G4gMuvlJSb0}, and the source code can be found at \url{https://github.com/soarsmu/IncBL}.
\end{abstract}

\begin{IEEEkeywords}
Bug Localization, Information Retrieval, Mining Software Repository
\end{IEEEkeywords}

\section{Introduction}
\label{sec:intro}

Information retrieval-based bug localization (IRBL) is a popular research topic in software engineering and has shown promising results in the last decade. The basic idea behind IRBL is to model the bug localization as an ad-hoc document search problem. Numerous research efforts have been invested in developing IRBL techniques~\cite{bugzbook, buglocator, BLUiR}. However, such research mainly aims to achieve better retrieval accuracy rather than make tools more efficient in software repositories where codebases evolve and new bug reports emerge. As software repositories are evolving, these tools need to re-process all files and update the entire model representations (e.g., tf-idf weights) to ensure accuracy, even for some minor changes (e.g., adding or deleting several lines of code), which limits their usages in time-sensitive or compute-intensive scenario.

To improve the efficiency of IRBL, Rao et al.~\cite{6671281} proposed an incremental update framework, but their approach cannot tackle some cases, e.g., when the number of documents changes\footnote{More detailed discussion can be found in our online appendix, that we make available at \url{https://github.com/soarsmu/IncBL/blob/main/discussion/appendix.pdf}}. Moreover, their method has only been applied to bug localization methods that purely rely on Bag-of-Words (BoW) models. Last but not least, there is no open-source artifact or tool that practitioners can adopt. The above facts motivate us to develop an incremental bug localization tool that addresses the limitations of Rao et al.'s work.

Akbar and Kak~\cite{bugzbook} divided IRBL tools that were published between 2004 and 2019 into three generations. The $1^{st}$ generation tools that are solely based on BoW models perform the worst, while $3^{rd}$ generation tools that utilize term-term order and semantics, e.g., SCOR \cite{SCOR}, require much time and computing resources to train models. Moreover, none of the tools specified as $3^{rd}$ generation in \cite{buglocator} is made publicly available. The $2^{nd}$ generation tools use structural information in codebases and software evolution information (e.g., historical bug reports) to improve accuracy. Two representatives of the $2^{nd}$ generation tools, BugLocator \cite{buglocator} and BLUiR \cite{BLUiR}, are evaluated in \cite{bugzbook}, and the empirical study shows that BugLocator outperforms BLUiR on Bugzbook \cite{buglocator}, a dataset containing bug reports from 27 large open-source repositories. Thus, we decide to build our incremental bug localization solution ({\em IncBL}) on top of BugLocator for the following reasons: It does not require as much time and computing resources as the $3^{rd}$ generation tools while performing better than many other $2^{nd}$ generation and the $1^{st}$ generation tools. Moreover, BugLocator is a popular tool and many other tools are built on top of it, e.g., the recently proposed Legion tool implemented in Adobe \cite{jarman2021legion}. Although this work focuses on making BugLocator incremental, the incremental design used in \toolname can also potentially be translated into other solutions that are based on BugLocator or have similar features to it.

\toolname performs a complete computation and retains the computed information, e.g., \textit{term frequency}, \textit{document frequency}, and other model parameters when it is applied on the underlying repository for the first time. Once a new bug report is raised, \toolname only updates the corresponding information, e.g., tf-idf weights, incrementally rather than re-compute the entire model, which can significantly reduce redundant computation and retrieval latency. We evaluate \toolname on Bugzbook dataset. Our evaluation results show that the \toolname can run 4.5 times faster (i.e., reducing  77.79\% of the running time on average) than the original BugLocator, without sacrificing the accuracy. 

To promote the usage of IRBL in practices and alleviates debugging costs for developers, we implement \toolname as a GitHub App, which can be installed in GitHub to locate potential potential buggy files for issues tagged as `bug' or be deployed locally for the same functionality.

The rest of this paper is organized as follows. Section~\ref{sec:buglocator} introduces the retrieval model and workflow of BugLocator. Section~\ref{sec:IncBL} describes the design and implementation of main features, and usage scenario of~\textit{IncBL}. Section~\ref{sec:eval} reports the evaluation results of \toolname on the Bugzbook dataset.  After surveying the related work in Section~\ref{sec:rel_work}, we conclude the paper and present future work in Section~\ref{sec:conclusion}.

\section{B\lowercase{ug}L\lowercase{ocator}}
\label{sec:buglocator}
\toolname extends BugLocator \cite{buglocator} with the support for incremental computing and integration into the GitHub platform. In this section, we briefly introduce the workflow of BugLocator and the information retrieval model used.

\vspace{0.2cm} \noindent \textbf{Step 1. Processing code files.} 
BugLocator preprocesses each Java code file in the codebase as follows. First, it utilizes a Java parser to extract identifiers (e.g., package name, method name, etc.) and then appends them to the original code contents. After that, BugLocator performs stemming and stopwords removal to produce a code corpus. A Vector Space Model (VSM) (more details of the incremental version of VSM will be given in Section \ref{subsec:design}) is then used to vectorize each document in the code corpus so that further steps (e.g., computing similarity) can be performed.


\vspace{0.2cm} \noindent \textbf{Step 2. Processing bug reports.} 
BugLocator first combines bug report titles with descriptions. Then, a bug corpus is created after performing stemming and stopwords removal on the combined documents. An important feature of BugLocator is to leverage bug-fixing history information to help rank faulty files, by computing the similarity between a bug report and all the past fixed bug reports. The intuition is that the current new bug report and the similar past fixed bug reports are likely to share the same source code files to be modified. BugLocator uses the VSM to compute similarities between bug reports. For a bug report $B$ and the $m^{th}$ code file referred by at least one bug report, their $SimiScore$ is computed by:
\begin{equation}
    SimiScore = \sum_{B' \in l(m)}\frac{Similarity(B,B')}{|l(m)|}
    \label{math:simiscore}
\end{equation}
where $l(m)$ is all the bug reports linked to the $m^{th}$ code file and $Similarity(B,B')$ is the similarity between $B$ and $B'$ computed using VSM. For a bug report and a code file that is not linked to any past bug report, their $SimiScore$ is $0$.

\vspace{0.2cm} \noindent \textbf{Step 3. Localizing buggy files.} 
BugLocator defines and computes the relevance score between bug reports and source code files. A VSM requires documents and queries to share the same vocabulary set, so BugLocator discards all the terms that appear in bug reports but not in the code corpus and then use the VSM produced in Step 1 to compute the $VSMScore$ between a document (source code file) $d$ and a query (bug report) $q$. Besides, BugLocator revises the VSM model by weighting $VSMScore$ with a function $g$ to favors long documents during ranking. BugLocator defines $g$ as
\begin{equation}
    g = \frac{1}{1 + e^{-N(|terms|)}} \text{, where}~N(x) = \frac{x - x_{min}}{x_{max} - x_{min}}
    \label{math:g}
\end{equation}
The $|terms|$ is the number of terms in preprocessed code files and $N(x)$ is a max-min normalization function. The final relevance score between the bug reports and source code files is a linear combination of revised $VSMScore$ and $SimiScore$:
\begin{equation}
    relevance = \alpha \times g \times VSMScore + (1-\alpha) \times SimiScore
\end{equation}
where $0\leq \alpha \leq 1$, and the performance is best empirically \cite{buglocator} when $\alpha$ is between $0.2$ and $0.3$.

\section{ Design and Usage Case of I\lowercase{nc}BL}
\label{sec:IncBL}
In this section, we introduce how the VSM used in BugLocator can be updated incrementally and how \toolname incorporates incremental computing in the tool. We also present how users can utilize \toolname as a GitHub App and a locally-deployed tool.

\subsection{Incremental Design}
\label{subsec:design}

VSM is used to represent a collection of documents (the code corpus and bug report corpus in this paper). First, we need to create $V$, which is the vocabulary of terms appearing in documents, and a term-document matrix $A$, whose size is $M \times |V|$, where $M$ is the number of documents. $A_m(w)$ represents the occurrence count of the $w^{th}$ term in $V$ in the $m^{th}$ file. In a VSM, each document is represented as a vector with the size of $|V|$. The $w^{th}$ value of this vector is the tf-idf weight that is computed by $tf_m(w) \times idf(w)$. The {\em term frequency} $tf_m(w)$ has different definitions over $A_m(w)$, and BugLocator computes it by $tf_m(w) = log(A_m(w)) + 1$. The $idf(w)$, called {\em inverse document frequency}, is computed by $idf(w) = log(\frac{M}{df(w) + 1})$ where $df(w)$ refers to {\em document frequency}, total number of documents that have the $w^{th}$ term. We concatenate representations for each source file into a big matrix, denoted by $D$, in which $D_m(w)$ presents the tf-idf weight of the $w^{th}$ term in the $m^{th}$ file. 
Note that an element $D_m(w)$ is computed with $(log(A_m(w)) + 1) \times \frac{M}{df(w) + 1}$, meaning that $D_m(w)$ needs to be updated only when at least one of the three values (i.e., $A_m(w)$, $df(w)$, and $M$) changes.

Next, we explain how to incrementally update $D$ when documents change.
At the document level, changes can be categorized into one or a combination of the following three atomic changes: (1) adding a new document, (2) deleting an existing document, (3) modifying an existing document, which corresponds to adding, deleting, and modifying a row in $A$. At the term level, the atomic changes are: (1) adding a new term, (2) deleting an existing term, and (3) changing the term frequency of an existing term in a document.

Let us consider two cases: 1) $M$ does not change and 2) $M$ changes. For the case in which $M$ does not change, only $df$ and $A$ have an influence on $D$. We first elaborate on how the term-document matrix $A$ can be updated. The update is straightforward. We first update the columns (term-level) and then update the rows (document-level) in $A$. If the $w^{th}$ term is deleted, we delete the $w^{th}$ column in $A$. If a new term is added, we append a new column to $A$ and set all the values as $0$. If the term frequency value of the $w^{th}$ term in $m^{th}$ document changes, we just need to set the corresponding position $A_m(w)$ to a new number. Then, we can update rows in a similar way. if the $m^{th}$ document is deleted, we delete the $m^{th}$ row in $A$. If a new document is added, we append a new row to $A$ and set this row as the term frequency of this newly added document. When the $m^{th}$ document is changed, the $m^{th}$ row in $A$ should be updated according to the modified term frequency of the document. We next elaborate how document frequency vector should be updated. Assuming the $m^{th}$ file is affected, the $df(w)$ can be updated with the following formula:
\begin{equation}
	df^{new}(w) = df^{old}(w)+ [sign(A_{m}^{new}(w)) - sign(A_{m}^{old}(w))]
\end{equation}
where $sign(\cdot)$ returns $+1$ for positive inputs, $-1$ for negative inputs, and 0 for 0. $A^{old}$ and $A^{new}$ are term-document matrices before and after this change. If the file $m$ or term $w$ are not in $A^{new}$ (or $A^{old}$), we set $A_{m}^{new}(w)$ (or $A_{m}^{old}(w)$) as $0$. 

\begin{figure}[t!]
	\centering
	\includegraphics[width=0.9\linewidth]{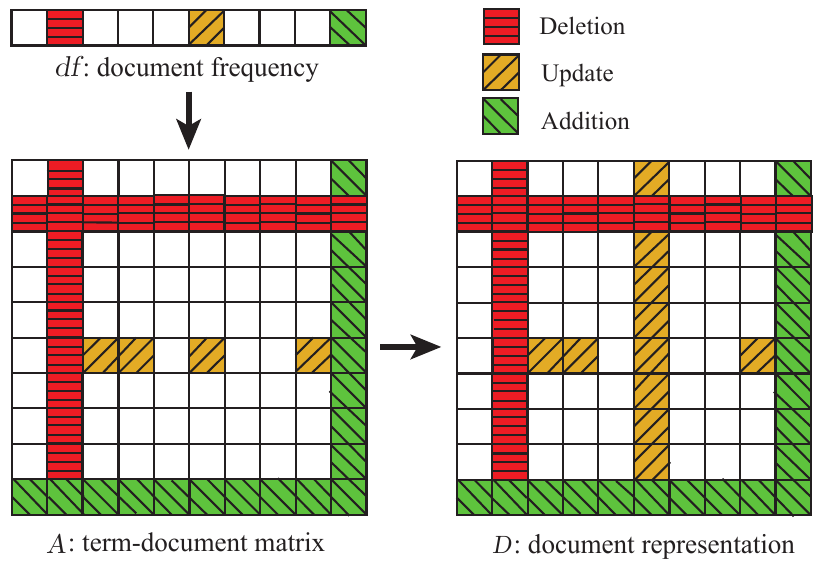}
	\caption{This figure illustrates how term-document matrix ($A$), document frequency ($df$) and document representation ($D$) can be updated incrementally. Each row in $D$ is the element-wise product of $df$ and the corresponding row in $A$, i,e, $df \odot A = D $. Blocks highlighted in red, green, and orange indicates that the corresponding positions in the matrices should be deleted, added, or modified, respectively. Colors in $df$ and $A$ will propagate to $D$, indicating that only the labeled blocks in $D$ need to be updated.}
	\label{fig:usage1}
\end{figure}

We illustrate how the term-document matrix $A$, the document frequency vector $df$, and document representation matrix $D$ can be updated according to the changes of documents and terms in Fig.~\ref{fig:usage1}. We label the columns in $A$ as green and red to represent term addition and deletion respectively. We highlight the rows in $A$ as green and red to represent document addition and deletion. Some blocks are highlighted with orange, which means that term frequency values at the corresponding positions are updated. We label the changes in $df$ using the same color scheme. $D$ is generated by an element-wise production between $df$ and $A$ (i.e., $df \odot A = D $), therefore in $D$, only the blocks that are produced by the changed portion in $df$ and $A$ need updates.  

So far, we have explained how term-document matrix $A$ and document frequency vector $df$ can be updated incrementally. Considering the changes between two versions are small (several commits), most elements in $A$ and $df$ will remain the same. As a result, when $M$ does not change, we can update $D$ incrementally at a very low cost: only positions highlighted with a color need to be updated.

However, if $M$ changes, the whole $idf$ vector needs updates, meaning that we must update all the values in $idf$, and all model parameters must be recomputed. We take this case in consideration as follows. We have $idf(w) = log(\frac{M}{df(w) + 1})$, which can be transformed to $log(M) - log(df(w) + 1)$. We can update $idf(w)$ by 
\begin{equation}
	idf^{new}(w) = idf^{old}(w) + log(\frac{M+\Delta M}{M})
\end{equation}
where $\Delta M$ is the change on M. In other words, we use the old $idf$ that is not affected to update the value of $idf$ incrementally rather than computing the $idf$ value from scratch.

Such changes are widespread in software development as codebases are continuously evolving and new bug reports are raised. We use this incremental model to update the VSMs in \textbf{Step 1} and \textbf{2} of the BugLocator as described in Section~\ref{sec:buglocator}. The updating method also indicates that \toolname only needs to preprocess the affected files rather than act like BugLocator who re-processes all files in the repositories each time it runs, which saves time even further.

\subsection{Implementation and Usage}
\label{subsec:implementation}
To promote the use of bug localization tools, we implement \toolname as a GitHub App. With some simple settings given in the tool homepage (\url{https://github.com/apps/incbl}), GitHub users can easily install \toolname in their public projects. Once \toolname is installed, it will automatically analyze the codebases and past bug reports. Each time when a new issue tagged with `bug' is raised, \toolname updates models incrementally and locates relevant buggy files for this issue. After files are retrieved, \toolname posts the top 10 most relevant files for the issue so developers can get notified. Please refer to our video (https://youtu.be/G4gMuvlJSb0) to see \toolname in action.
Users can also deploy \toolname locally on their own machines. Only a simple Python command is needed to run {\em IncBL}. Users just need to specify the path to the codebases under analysis, the path to bug reports, and the path for storage.
\section{Evaluation}
\label{sec:eval}

\subsection{Experimental setting}\label{subsec:dataset}
We use Bugzbook~\cite{bugzbook} as our benchmark to evaluate {\em IncBL}. Bugzbook contains releases of 27 software projects and bug reports that correspond to each release. Following Akbar and Kak~\cite{bugzbook}, we focus on Java, Python, C, and C++ code files, and discard other files in the codebase as well as bug reports linked to other file types in this experiment. We end up with 43,017 bug reports.

We aim to examine if \toolname can run more efficiently without losing any accuracy. For this purpose, we measure and compare the running time of \toolname and re-computing from scratch. To measure the running time of {\em IncBL}, we consider the usage scenario mentioned in Section \ref{subsec:implementation}: a user raises an issue in GitHub and then \toolname updates the VSMs incrementally to localize buggy files with lower latency. We measure the running time that \toolname requires for locating potential buggy files for the raised bug report, denoted as $T_{Inc}$. We apply BugLocator on the usage scenario and measure its running time for each bug report, denoted as $T_{BL}$. Conducting the experiments on all collected bug reports is very time-consuming since BugLocator needs to compute everything from scratch. Therefore, we randomly sample 381 reports from all the studied bug reports, and this forms a statistically representative sample considering a 95\% confidence level and 5\% interval. We measure the running of \toolname and BugLocator on these sampled 381 reports. 
When measuring the accuracy, we keep the same setting as Bugzbook experiment: associating a group of bug reports with one software release. We run \toolname on all the 43,017 bug reports and incrementally updates VSM representations for each new software release. 


We run our experiments on a server with a 16-core CPU and 512 GB memory. The dataset, experimental results, and detailed instructions to run the experiments can be accessed via \url{https://github.com/soarsmu/IncBL}.

\begin{figure}[t!]
	\centering
	\includegraphics[width=0.85\linewidth]{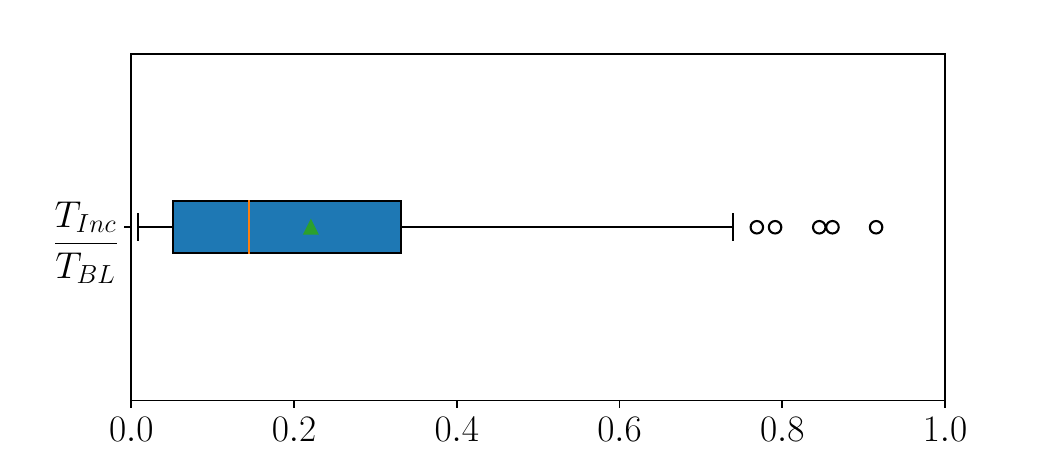}
	\caption{Boxplot of the ratio of the running time to perform bug localization on all studied bug reports of \toolname over that of the original BugLocator ($\frac{T_{Inc}}{T_{BL}}$). The median and mean are 14.44\% and 22.21\%, corresponding to 6.30 and 4.50 times speed up.}
	\label{fig:RQ1}
\end{figure}

\subsection{Results}\label{subse:results}
\textbf{\toolname can significantly reduce the running time for locating bugs in continuously evolving repositories by 77.79\% (i.e., 4.5 times faster) on average compared with the original BugLocator while maintaining the same accuracy.}
Fig.~\ref{fig:RQ1} presents the boxplot of the ratio of running time between \toolname and BugLocator.
The median and mean ratios are 14.44\% and 22.21\%. Wilcoxon signed-rank test shows that the reduction in running time is statistically significant ($p$-value $<$ 0.01). The amount of time saved by the \toolname is mainly related to two factors: (1) the number of affected files and (2) the number of existing files. The more files change, the more updating operations are needed. Besides, if there are too many existing files, the VSM can be very large and \toolname needs more time to identify the correct locations in matrices to perform updates. On the Bugzbook dataset, \toolname achieves a Mean Average Precision (MAP) of 0.331 -- on average, the correct files appear in the top 3 locations.


\section{Related Work}
\label{sec:rel_work}

Although many efforts have been put into improving the effectiveness of bug localization techniques~\cite{bugzbook, buglocator, BLUiR}, the long running time of such tools limits their real-world usages. Therefore, improvement of the efficiency of bug localization tools is needed. However, there is no other studies in the research community apart from Rao et al.'s framework in \cite{6671281} and \cite{rao2015comparing}. The limitations of \cite{6671281} were discussed in Section \ref{sec:intro}, and are addressed by \emph{IncBL}. \cite{rao2015comparing} only considers incremental LSA and LSA has been shown to perform worse than VSM in bug localization \cite{Rao2011}. Moreover, \toolname is the first incremental bug localization tool integrated as a GitHub App.
BugLocalizer \cite{thung2014buglocalizer} is a tool that implements BugLocator as a Bugzilla plugin. However, it does not support incremental updates.

\section{Conclusion and Future Work}
\label{sec:conclusion}
To help developers localize bugs, this paper presents {\em IncBL}, which can update the VSM incrementally and avoid repetitive computation, on top of BugLocator. Our evaluation shows that on average \toolname can run 4.5 times faster than re-computing the model from scratch while maintaining the same level of accuracy. We implement \toolname as a GitHub App that can be easily installed to analyze public repositories on Github. When a new bug report is raised, \toolname will update the VSM incrementally and notify developers about the potentially buggy files. Users can also deploy \toolname on their own machine to analyze repositories locally. In the future, we plan to create variants of \emph{IncBL} built on top of other bug localization tools.

\section*{Acknowledgment}

This research was supported by the Singapore Ministry of Education Academic Research Fund (AcRF) Tier 1 grant.

\bibliographystyle{IEEEtran}
\bibliography{reference}

\end{document}